\documentclass{svproc}
\usepackage{graphicx}
\usepackage{footmisc}
\usepackage{amsmath}
\usepackage [autostyle]{csquotes}
\usepackage{algorithm}
\usepackage[noend]{algpseudocode}

\usepackage[square,sort,comma,numbers,sectionbib]{natbib} 
\usepackage{url}

\begin{document}
\mainmatter              
\title{\large Local Sharing and Sociality Effects on Wealth Inequality in a Simple Artificial Society }
\titlerunning{Local Sharing}  
%
\author{John C. Stevenson }
\authorrunning{JC Stevenson} 
%
%
\institute{Long Beach Institute, Long Beach, NY 11561\\
\email{jcs@alumni.caltech.edu}
}

\maketitle              

\begin{abstract}
	
	Redistribution of resources within a group as a method to reduce wealth inequality is a current area of debate.  The evolutionary path to or away from wealth sharing is also a subject of active research. In order to investigate effects and evolution of wealth sharing, societies are simulated using a  minimal model of a complex adapting system.  These simulations demonstrate, for this artificial foraging society, that local sharing of resources reduces the economy's total wealth and increases wealth inequality. Evolutionary pressures strongly select against local sharing, whether globally or within a individual's clan, and select for asocial behaviors. By holding constant the gene for sharing resources among neighbors, from rich to poor, either with everyone or only within members of the same clan, social behavior is selected but total wealth and mean age are substantially reduced relative to non-sharing societies. The Gini coefficient is shown to be ineffective in measuring these changes in total wealth and wealth distributions, and, therefore, individual well-being. Only with sociality do strategies emerge that allow sharing clans to exclude or coexist with non-sharing clans. These strategies are based on spatial effects, emphasizing the importance of modeling movement mediated community assembly and coexistence as well as sociality.
	
\end{abstract}
\section{Introduction}

Redistribution of wealth is often proposed as a method to reduce wealth inequality. However, the actual effect of wealth sharing on inequality is a subject of debate among economists and policymakers \citep{davies1986does,glaeser2005inequality,bossmann2007bequests}. Concurrent with this debate are questions on the evolutionary stability of an social system with sharing norms or policies. This stability is important for long term viability, resiliency, social cooperation, and trust \cite{reggiani2002resilience,ward,wilson1971}. Evolutionary explanations of emergence of wealth inequality from small scale, relatively egalitarian societies that are found in both archaeological and ethnographic data are active areas of research \cite{mattison2016evolution}.

A spatiotemporal agent based model with local sharing and sociality behaviors within and among multiple groups (\textquote{clans}) provides a fresh perspective on both the effectiveness of wealth sharing and the evolutionary origins and stability of societies with these behaviors. This model, as a minimal model of a system \citep{roughgarden}, has the dynamics of a complex adaptive system (CAS) and allows intragroup (CAS2) and intergroup (CAS1) evolutionary optimization \citep{wilsonCAS}. Altruistic behavior, like that of local sharing, may require optimization at the intergroup (CAS1) level \citep{axelrod,howard,pepper,wilson,wilsonCAS}. With the techniques of modern coexistence theory \citep{chessMech}, the ability of sharing clans to withstand, invade, exclude, and even coexist with selfish and cheating clans is explored. These strategies are dependent on, and reflect the importance of, sociality preferences as well as movement mediated assemblies of ecologies and societies \cite{schlagel} 

This paper begins with a description of the modeling of an artificial foraging society, linking this particular simulation to standard mathematical models of biology and ecology. The implementation and selective pressures of sociality and local sharing for both a single population and a population divided into agent recognizable clans are described in detail. The fitness of populations with various alleles of these genes is measured through evolutionary optimization. Zones of coexistence and exclusion for a sharing clan paired with a non-sharing clan are identified and discussed. 

\section{Methods}
The agent based model is described and the parameters of the agents and the landscape are detailed in sufficient detail to allow reproduction of these results. The resultant population dynamics are referenced to the standard biological, ecological, and stochastic gene-frequency mathematical models. The sociality and local sharing behaviors for a single group are described and the extension for multigroup sociality is detailed. The results of evolutionary optimization of single and multigroup populations with randomly mixed sociality and local sharing behaviors are presented and discussed. The coexistence and exclusions resulting from a sharing clan interacting with a selfish clan as functions of sociality are presented.

\subsection{Basic Agent Based Model}
This spatiotemporal, multi-agent-based model (ABM) is a minimum model of a system \citep{roughgarden} based on Epstein and Axtell's classic Sugarscape \citep{axtell,stevenson}. Table \ref{table:appA} provides the definition of the agents' and landscape's parameters used for this investigation. Vision and movement are along rows and columns only. The neighborhood is von Neumann, four cells each one step away. The two dimensional landscape wraps around the edges (often likened to a torus). Agents are selected for action in random order each cycle as shown in Algorithm \ref{action}. Birth cost is the parent's required surplus resources that are consumed during reproduction. Given that the birth cost and free space constraints are met, the probability of reproduction, $p_{f}$, is expressed as infertility $ f = \frac{1}{p_{f}}$. Reproduction is haploid (cloning) with a stochastic, single-point mutation of an agent's genome. The agents interact on an equal opportunity (flat) landscape of renewing resources. The landscape parameters remain constant during a run. With this approach, no endowments for the newborn are required whether for new births or at start-up. Once all the agents have cycled through, the landscape replenishes at the growth rate and the cycle ends.  Agents only die when their metabolism exhausts their current resources, otherwise they are immortal. The extensions of this model for local sharing and sociality are discussed in detail in the following sections.

The population dynamics that emerge from this simple underlying model have been shown to agree with discrete Hutchinson-Wright time delayed logistic growth models for a single species of mathematical biology and ecology \citep{murray,stevenson,hutch1,kot,stevensonEcon}, with standard Wright-Fisher class, discrete, stochastic, gene-frequency models of mathematical population genetics for finite populations, \citep{ewens,moran,cannings,stevenson}, and with modern coexistence theory for multiple species \citep{chessMech,stevensonX}.

\begin{table}[h!]
  \begin{center}
	\begin{tabular}{|c|c|c|c|c|}
		\hline
		Agent Characteristic & Notation & Value & Units & Purpose \\
		\hline
		vision & $v$ &  6 &  cells & vision of resources on landscape \\
		movement & -- &  6 &  cells per cycle &  movement about landscape \\
		metabolism & $m$ & 3 & resources per cycle &  consumption of resource \\
		birth cost & $bc$ & 0 & resources &  sunk cost for reproduction \\
		infertility & $f$ & 1-85 & 1/probability & likelihood of birth \\
		puberty & $p$ & 1 &  cycles &  age to start reproduction\\
		surplus & $S$ & 0+ & resources &  storage of resource across cycles \\
		mutation & $\mu$ & $>=0$ & probability & mutation rate \\
		group identity & $ID$ & 0-9 & allele & clan identification \\
		same sociality & $ixS$ & 0-2 & allele & see Table \ref{table:singleS} \\
		other sociality & $ixO$ & 0-2 & allele & see Table \ref{table:singleS} \\
		local sharing & $ix$ & 0-2 & allele & see Tables \ref{table:singleS} and \ref{table:twoS}  \\
			\hline
		\end{tabular}
		\bigbreak
		\begin{tabular}{|c|c|c|c|}
			\hline
			Landscape Characteristic & Notation & Value & Units\\
			\hline
			rows & $r$ & 50 & cells \\
			columns & $c$ & 50 & cells\\
			max capacity &$R$ & 4 & resource per cell\\
			growth & $g$ & 1 & resource per cycle per cell \\
			initial & $R_{0}$ & 4 & resource, all cells\\
			\hline
		\end{tabular}
		\caption{Agent (Top) and Landscape (Bottom) Parameters of the ABM}
		\label{table:appA}
	\end{center}
\end{table}


\subsection{Single Group Sociality}

\begin{table}[h!]
	\begin{center}
		\begin{tabular}{|c|c|c|}
			\hline
			Social Gene & Label & Behavior \\
			\hline
			$0$ & asocial & avoids everyone\\
			$1$ & social & seeks everyone\\
			$2$ & neutral & no preference\\
			\hline
		\end{tabular}
		\begin{tabular}{|c|c|c|}
			\hline
			Share Gene & Label & Shares With \\
			\hline
			$0$ & selfish & no one\\
			$1$ & generous & everyone\\
			$2$ & selective & own clan\\
			\hline
		\end{tabular}
		\caption{Sociality (Left) and Local Sharing (Right) Alleles for a Single Group}
		\label{table:singleS}
	\end{center}
\end{table}
\begin{algorithm}
	\begin{algorithmic}[1]
		\small
		\Procedure{action cycle}{}
		\State set action list to all live agents, done list to $NULL$ 
		\While{action list not empty}
		\State select random agent $a_{i}$
		\State find target cell $T({a_{i}})$ with closest best resource within vision movement $v$
		\If{$j$ ties for $T^j(a_{i})$}
		\State $T^{best} = \Call{scoreSocial}{T^j(a_{i})$} on sociality
		\EndIf
		\State forage best cell resource at $T^{best}$
		\State metabolize $m$ resources
		\If{ $a_{i}$ surplus  $S(a_{i})<0$}
		\State best cell $bJ = \Call{shareRequest}{a_{i}}$
		\If{ $S(a_{i})+(S(a_{bJ})-m)<0$}
		\State $a_{i}$ dead ,remove from action and done lists
		\Else
		\State shared resource $R_{S} = m-S(a_{i}) $ 
		\State $S(a_{i}) += R_{S}$ and $S(a_{bJ}) -= R_{S}$
		\EndIf
		\EndIf
		\If{prob $f$ \&  $a_{i}$ age $>p$ \& free space \& $S(a_{i})> bc$}
		\State reproduce into random free neighbor space
		\State $S(a_{i})-= bc$ 
		\If{p==0}
		\State add child to action list
		\Else \State add child to done list
		\EndIf
		\EndIf
		\State remove $a_{i}$ from action list, add to done list
		\EndWhile
		\EndProcedure
		\Procedure{shareRequest}{$a_{i}$}
		\State best available resource $bR=0$ and best neighbor $bJ=NULL$
		\For{ each von Nueman neighbor $a_{j}$ }
		\If{$a_{j}$ will share with $a_{i}$ and has resouces $S_{j}> m$}
		\If{$S_{j}-m > bR$}
		\State $bJ=j$
		\EndIf
		\EndIf
		\EndFor
		\State \Return $bJ$
		\EndProcedure
		\Procedure{scoreSocial}{$T_{a_i}^j$}
		\For {each target cell von Neiman neighbor $j$ }
		\State $score_{j}=0$
		\State count clan neighbors $cN$ and other neighbors $oN$
		\If{ $a_{i}$ is clan social/(asocial)}
		\State $score_{j} +/(-)= cN$
		\EndIf
		\If{ $a_{i}$ is other social/(asocial)}
		\State $score_{j} +/(-)= oN$
		\EndIf
		\EndFor
		\State \Return $j$ with best $score_{j}$
		\EndProcedure
	\end{algorithmic}
	\caption{\small Action cycle for ABM}
	\label{action}
\end{algorithm}

\begin{figure}
	\begin{center}
		\resizebox{\columnwidth}{!}{
			\includegraphics[angle=-90,scale=1.0]{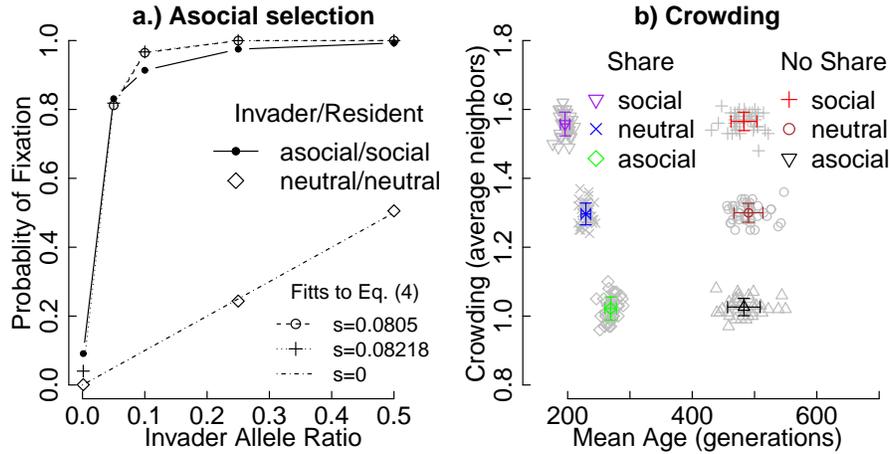}
		}
	\end{center}	
	\caption {Sociality Selection Pressure and Resultant Crowding}
	\label{fig:weakCrowd}
\end{figure}

A preference is included for the occasions when both the resource amount and shortest distance to the best available foraging resource is a tie between two or more cells. This preference is represented by a gene with the alleles described in Table \ref{table:singleS} (Left) and implemented as described in Algorithm \ref{action}. The selective pressure exerted for this preference can be quantified as follows\citep{ewens}. For the Wright-Fisher class of models with no mutations or selective pressure, the probability $\pi$ of one of the two alleles $(A_{1},A_{2})$ overtaking the whole population (fixation) is given as:
\begin{equation}
	\pi=i/2N
\end{equation}
where $i$ is the initial population of one of the two alleles and the total population is $2N$. The selection of $2N$ for the haploid population allows comparison with diploid populations of standard Wright-Fischer models. Using the Moran model \cite{ewens}, the probability that the next individual selected to die is an $A_{1}$ allele is 
\begin{equation}
	\mu_{1}i/\{\mu_{1}+\mu_{2}(2N-i)\}
\end{equation}
where $i$ is the number of $A_{1}$ alleles in the $2N$ population and $\frac{\mu_{1}}{\mu_{2}}$ defines the selective advantage $s$. If $\mu_{1}=\mu_{2}$ there is no selective advantage whereas if $\mu_{1}<\mu_{2}$ then allele $A_{1}$ has a small selective advantage. The probability of fixation from Equation(1) is now:
\begin{equation}
	\pi_{i} = \{1-(\mu_{1}/\mu_{2})^{i}\}/\{1-(\mu_{1}/\mu_{2})^{2N}\}
\end{equation}
By defining $\mu_{1}/\mu_{2} = 1 - s/2$ with $s$ small and positive, Equation (3) can be approximated as
\begin{equation}
	\pi(x) = \{1-\exp(-\alpha x/2)\}/ \{1-\exp(-\alpha/2)\}
\end{equation}
where $x=i/2N$ and $\alpha = 2Ns$. With this approximation, the magnitude of the selection pressure as represented by $s$ for allele A1 can be implied as an invading species and A2 a resident species with a initial A1 allele ratio defined as $i/2N$ as shown in Figure \ref{fig:weakCrowd}a. While $s=0.08$ is a somewhat weak selection pressure due to the relatively rare opportunity for preference, the selection pressure has significant macroscopic effects as shown by Figure \ref{fig:weakCrowd}b with an 5\% initial allele ratio having an 80\% probability of fixation.

\subsection{Local Sharing}

In addition, the ability to share resources under duress is controlled by second gene with alleles described in Table \ref{table:singleS} (Right)\footnote{For single group simulations, the generous and selective alleles result in identical behavior. A clan with a selfish allele is also called cheater when paired with a generous clan for clarity.}. Specifically, at the end of an agent's action cycle, if their resources are so depleted that they face imminent death, they will ask their neighbors for sufficient resources to survive the action cycle. All agents will make such a request but only neighbors with allele generous or selective may be willing to share. Of the four neighbors, the neighbor with the most surplus resources greater than their metabolism rate and willing to share will provide resources up to the amount needed by the requesting agent (see Algorithm \ref{action}).

Evolution of the sociality and sharing provides relative measures of intragroup fitness. If all possible alleles for both the sociality gene and the sharing gene are randomly assigned to the population, generous and selective sharing alleles are excluded; and the asocial sociality allele excludes neutral and social alleles. If the sharing gene is fixed as generous, then social will exclude neutral and asocial sociality alleles. In all these evolutionary selections, the optimization is based on the individual (CAS2).

\subsection{Multiple Group Sociality and Sharing}

The $ID$ gene allows agents to identify other agents within their vision as either of the same clan or \textquote{other} clans. A second sociality gene is added such that social behavior is now governed by two genes, one ($ixS$) for interaction with agents with the same ID (a fellow clan member) and a separate gene ($ixO$) for agents with a different ID (other clans). Table \ref{table:twoS} provides the nine possible social personalities with their allele values, resultant behavior, and labeling. Sharing adds the "selective" allele to the sharing regimes as described in Table \ref{table:singleS}. Of particular interest is the pairing of a generous clan with a selfish clan. The selfish group in this situation is often referred to as a cheater since they will request and receive resources from the generous clan but they will not reciprocate. The selective sharing clan is immune to cheaters.

\begin{table}[h!]
	\begin{center}
		\begin{tabular}{|c|c|c|c|}
			\hline
			Clan Social Gene & Other Social Gene & Label & Behavior\\
			\hline
			$0$ & $0$ & loner & avoids everyone\\
			$1$ & $0$ & clannish & seeks clan,avoids others\\
			$2$ & $0$ & shy & avoids others\\
			$0$ & $1$ & aggressive & avoids clan, seeks others\\
			$1$ & $1$ & gregarious & seeks everyone\\
			$2$ & $1$ & friendly & seeks others\\
			$0$ & $2$ & outcast & avoids clan\\	
			$1$ & $2$ & homebody & seeks clan\\
			$2$ & $2$ & neutral & no preferences\\
			\hline
		\end{tabular}
		\caption{Sociality Alleles for Own Group and Other Groups Genes}
		\label{table:twoS}
	\end{center}
\end{table}

\begin{figure}
	\begin{center}
			\includegraphics[angle=-90,scale=0.5]{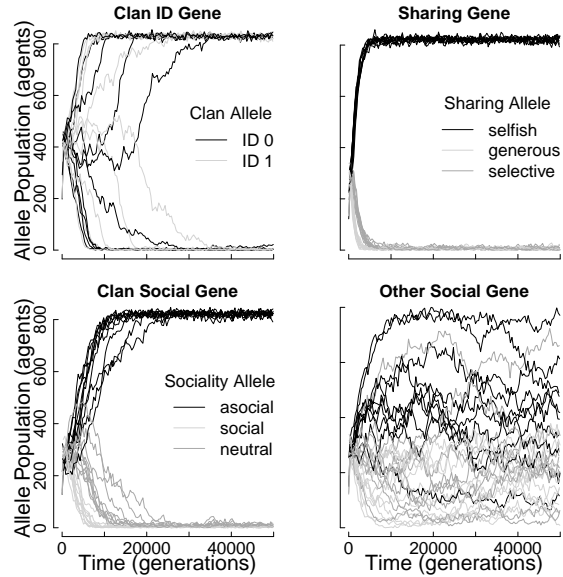}
	\end{center}	
	\caption {Evolution of sharing and sociality alleles}
	\label{fig:shareNoEO}
\end{figure}

Evolution of sociality and sharing genes for pairs of clans provides relative fitness measurements of intergroup fitness, suggesting the possibility of clan level optimization (CAS1). Two clans of equal size are generated with the total population half the expected carry capacity of the landscape\footnote{The expected carry capacity is the number of agents $N_{a}$ whose metabolism $m$ consumes the maximum amount of resources ($g*r*c$) that can be replenished in one action cycle. That is $N_{a} = grc/m$.}. By allowing the agents to initially grow into a somewhat empty landscape, the opportunity to acquire surplus resources is provided. When all possible alleles for both the two sociality genes and the sharing gene are randomly assigned to both group populations and the simulation is run to a steady state species population, both generous and selective sharing alleles are excluded as well as neutral and social sociality alleles. Figure \ref{fig:shareNoEO} shows these exclusions as well as the feedback that occurs when one of the groups stochastically gains a small population advantage. At that point, the clan asocial allele influence drives the larger clan to surround the smaller clan, quickly leading to the exclusion of the less populous clan. Once one clan has been eliminated, the other sociality gene has no effect and displays stochastic diffusion of its alleles. All these results are strongly suggestive of CAS2 optimization, perhaps because there is not sufficient time or initial separation to allow for CAS1 strategies to emerge.

\subsection{Coexistence and Exclusion}
In addition to the evolutionary results of one and two groups of agents with initial populations of randomly seeded alleles for sociality and sharing, two populations with fixed and different sociality and sharing alleles (different species) were studied. These simulations determine whether a particular species can survive under the inevitable mutation of one of these alleles or from immigration from outside the landscape. Based on the demonstrated evolutionary pressure against sharing alleles, the question is whether a given sharing species is capable of surviving an invasion of a small population of a selfish species. Modern coexistence theory answers this question by having a small  population of an invading species invade a resident population \citep{chessMech}. Three outcomes are possible, the invader fails and is excluded, the invader succeeds and excludes the resident species, or both species coexist. For definitive coexistence, the roles of resident and invading species must then be reversed with the same coexistence result over multiple runs. The two sharing species were individually tested against a selfish species. The invading population was set at $5\%$ of the resident population with the total of both populations initially set at half the carry capacity as before.

\section{Results}
The total wealth and wealth inequality distributions for single and multiple groups with specific sharing and sociality alleles are generated and compared. Results for evolutionary selection on a single group with equally distributed sharing alleles, equally distributed sociality alleles, and both at once are reported and discussed. Relative fitness and zones of coexistence and exclusion are generated and discussed based on modern coexistence theory \citep{chessMech} and in terms of complex adaptive system optimization \citep{wilsonCAS}.

\subsection{Single Group Wealth Inequalities}
\begin{figure}
	\begin{center}
		\resizebox{\columnwidth}{!}{	
			\includegraphics[angle=-90,scale=1.0]{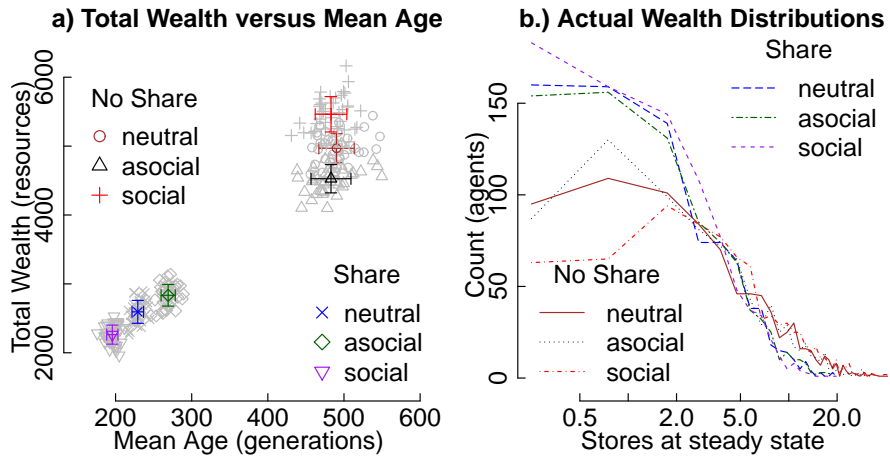}
		}
	\end{center}
	\small	
	\caption {Single Group Sharing and Sociality Effects on Total Wealth}
	\label{fig:oneGroup}
\end{figure}

Figure \ref{fig:oneGroup}a presents the total wealth of the population as a function of the mean age of the population for a single group. The effects of local sharing are dramatic and detrimental. The total wealth of the population and its mean age are both halved by sharing. Since the population maintains the steady state carry capacity, the death and birth rates (for which mean age is a good proxy) are much higher for local sharing. Figure \ref{fig:oneGroup}b provides the distribution of wealth among the steady state populations. These distributions confirm that a local sharing allele both increases the number of poor people and reduces the number of rich people relative to the selfish allele. Coupled with the halved total wealth of the population with sharing, there are no agents better off with the sharing allele. Surprisingly, the Gini coefficient \citep{fontanari2018gini,taleb2015not} was an ineffective measure of these wealth disparities as shown by Figure \ref{fig:heatCoex}a. Though evolutionary pressures at the individual level often prefer species that reproduce more quickly, die younger, and have little surplus resources \citep{stevensonEcon}, which results in a \textquote{tragedy of the commons} \citep{ostrom1990governing}, in this case, surprisingly, the selfish individual behavior appears to be better for the commons as well.

\subsection{Invasions and Cheaters}

A big concern for the survival of sharing clans is the presence of cheaters, who request stores when in need from neighbors but will not share with anyone when they receive a request. Successful requests result in only enough stores to survive the current action cycle. The possibility of deceit by the requester is beyond the scope of this research.

\begin{figure}
	\begin{center}
		\resizebox{\columnwidth}{!}{	
			\includegraphics[angle=-90,scale=1.0]{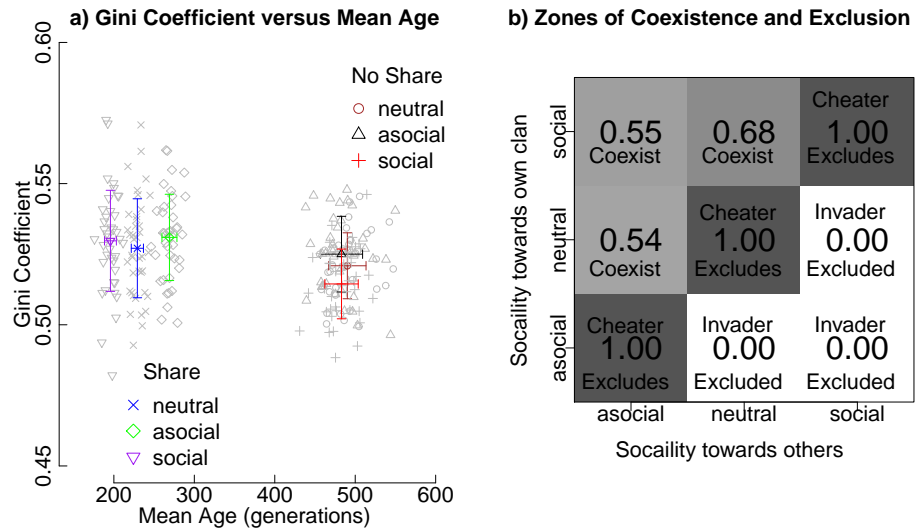}
		}
	\end{center}
	\small	
	\caption {a) Gini Coefficient by Mean Age b) Zones of Coexistence}
	\label{fig:heatCoex}
\end{figure}

\begin{figure}
	\begin{center}
		\resizebox{\columnwidth}{!}{
			\includegraphics[angle=-90,scale=1.0]{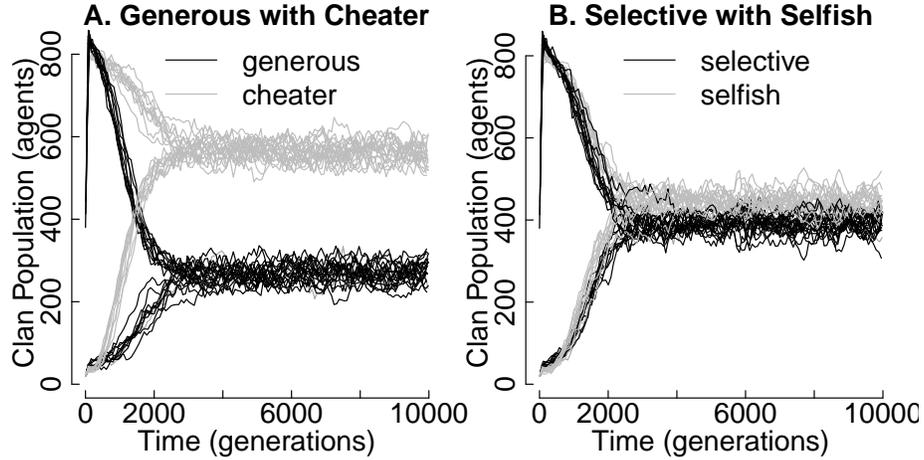}
		}
	\end{center}	
	\caption {Coexistence of Sharing alleles with Selfish allele}
	\label{fig:coXsh}
\end{figure}

When testing for possible coexistence zones for a pairing of a sharer clan with a selfish clan, each clan is tested as both invader and resident. A mutation is assumed to occur in the sharing gene resulting in a small population of invaders with the same sociality alleles but a different sharing allele. Figure \ref{fig:heatCoex}b identifies exclusion and coexistence zones as functions of both clans' sociality genes. Surprisingly, the results are broadly similar for both sharing clans: generous and selective. Three different zones emerge. One is the \textquote{Invader Excluded} zone where the resident species (of all three sharing alleles), with aggressive social behavior towards the invader, excludes the invader by depleting  the neighborhood of the less numerous invader of both space and resources.  A second zone emerges when the clan sociality allele is identical to the other sociality allele, thereby removing the social distinction of separate  clans. In this zone, the cheating clan excludes the sharing clans as both invader and resident. A third zone is one of coexistence with feedback fixing the selfish/cheater to total population ratio at approximately $1/2$ for most zones though the cheater/generous pair with homebody sociality is stable a higher cheater ratio. Figure \ref{fig:coXsh} presents these steady state clan population trajectories for 10 runs as invader and 10 runs as resident for each of the two representative pairings.

\section{Discussion}

The addition of sociality and local sharing functionality has increased the complexity of the foraging society's behaviors and, surprisingly, increased wealth inequity and reduced the total wealth of the society. In particular, the individual wealth distributions demonstrated that a clan with sharing alleles has more and poorer members as well as fewer and less rich members than a selfish clan, regardless of sociality preferences.  The sharing alleles were also shown to be less fit than the selfish allele and were consistently selected against within groups with both sharing and selfish alleles. There is no evidence of clan level optimization (CAS1). In spite of these results under evolutionary pressure, established clans with either generous or selective sharing alleles, under some sociality configurations, were able to coexist with selfish/cheating clans and, as residents in some cases, were capable of excluding the invading selfish/cheaters. The exclusion by a resident clan of an invading clan occurred because the sociality alleles promoted spatial attraction to the invaders, resulting in suffocation of the invading clan through lack of space and resources.  The coexistence of a sharing clan with a selfish/cheating clan was driven to a precise and stable ratio of clan populations, a result of feedback of spatial effects of social alleles on the availability of space and resources. The exclusion of sharing alleles by the selfish allele as both intruder and resident occurred when the sociality alleles were equal in both clans. This lack of sociality distinction between the clans allowed the selfish to dominate, demonstrating the importance of sociality for the survival of sharing clans. These exclusions of an invading clan, and coexistences of a sharing clan with a selfish clan, are dependent on the modeling of movement-mediated assembly of communities \citep{schlagel}.

%
%
\footnotesize
\bibliographystyle{spphys}

\bibliography{feb03_23.bib} 

%


\end{document}